\documentclass[prb,twocolumn,showpacs]{revtex4}   

\usepackage{graphicx}

\begin{document}

\title
{
Spin, charge and orbital fluctuations in a multi-orbital Mott insulator
}

\author
{                                                       
Akihisa Koga and Norio Kawakami
}

\affiliation
{
Department of Applied Physics, Osaka University, 
Suita, Osaka 565-0871, Japan
}

\author
{                                                       
T.M. Rice and Manfred Sigrist
}

\affiliation
{
Theoretische Physik, 
ETH-H\"onggerberg, 8093 Z\"urich, Switzerland
}
\date{\today}

\begin{abstract}
The two-orbital degenerate Hubbard model with distinct hopping 
integrals is studied 
by combining dynamical mean-field theory with quantum Monte Carlo simulations.
The role of orbital fluctuations for  
the nature of the Mott transition is elucidated by examining the temperature dependence
of  spin, charge and 
orbital susceptibilities as well as the one-particle spectral function.
We also consider the effect of the hybridization between the two
orbitals, which is important particularly close to  the Mott 
transition points. The introduction of the hybridization induces 
orbital fluctuations,
resulting in the formation of a Kondo-like heavy-fermion behavior, similarly to
$f$ electron systems, but involving electrons in bands of comparable width. 
\end{abstract}


\maketitle

\section{Introduction}\label{intro}
Strongly correlated electron systems with multi-orbital bands pose a
variety of intriguing problems.  One of the recently debated topics 
is the orbital-selective Mott transition (OSMT) in
highly correlated $d$-electron systems.
\cite{Anisimov,Fang,KogaLett,Sigrist,Liebsch,sces} 
It is a fundamental issue of multi-orbital systems whether 
Mott-transitions would take place in sequence or simultaneously for
all bands, if correlation would gradually be turned on. 
There are however also specific materials which have been discussed in this
context such as the calcium-doped single layer strontium ruthenate
$\rm Ca_{2-x}Sr_{x}RuO_4$\cite{Nakatsuji} and 
the ternary nickel oxide $\rm La_{n+1}Ni_nO_{3n+1}$,\cite{LaNiO,Kobayashi}
where the chemical substitution (or the change in the temperature) 
may trigger the OSMT in the $t_{2g}$ ($e_g$) orbitals
in the former (latter) case.

The extensive studies on the Mott transition in the multi-orbital systems
 clarified that  the competition between 
the intra- and inter-orbital interactions as well as the 
Hund coupling plays a key role to determine  the nature 
of the Mott transition.\cite{KogaLett}  It was found that under
special conditions in a two-band system the Mott-transitions
may merge to a single one, but would split for a generic form of the model. 
In particular, the presence of Hund coupling seems to be essential
to observe distinct transitions.  These conclusions 
were drawn from the analysis of the quasi-particle weights computed 
at zero temperature. In order to characterize the transitions the behavior of
the spin, charge and orbital fluctuations provides additional valuable
information. A systematic study of the temperature dependence of
certain susceptibilities will give us the necessary insight to analyze
in particular the electronic degrees of freedom which are localized 
through the Mott transition.

The above discussions on the Mott transition are restricted so far 
to systems for which the bands  do not 
hybridize, but are coupled to each other
only through electron-electron interactions.
However, the  hybridization between the bands may be important in some 
compounds.\cite{Kusunose}  In particular, this effect could give rise to a 
qualitative change in the phase diagram,  
when there occurs the OSMT, for which  
the intermediate phase appears with one orbital localized and 
the other itinerant. One thus naively wonders whether
Kondo-like  heavy fermion states would  be induced  by the hybridization 
between the orbitals. 
In fact, certain observed features can possibly be attributed to Kondo like behavior 
in the compound 
$\rm Ca_{2-x}Sr_xRuO_4$ $(0.2<x<0.5)$,\cite{Nakatsuji}
where the hybridization between orbitals
is induced by the tilting of RuO$_6$ octahedra.\cite{tilting} 
It is surprising that this behavior emerges from
electrons which originate from bands of comparable width.
These interesting observations naturally motivate us to
explore the effect of hybridization in more detail. 

In this paper, we study a two-orbital Hubbard model
with the distinct hopping integrals by combining dynamical mean field theory 
(DMFT)\cite{Metzner,Muller,Georges,Pruschke} 
with quantum Monte Carlo (QMC) simulations.\cite{Hirsch,Sakai}
We examine the spin, charge and orbital fluctuations which give insight into
the electronic properties in the regime of the OSMT.
We further consider the effect of hybridization, which may be 
important in real materials, and show that heavy-fermion-like behavior
emerges upon introduction of the hybridization.
The paper is organized as follows.
In \S\ref{sec2}, we introduce the model Hamiltonian for the two-orbital 
system and briefly explain the framework of DMFT. 
We discuss how the spin and orbital fluctuations affect 
the metal-insulator transition in \S \ref{sec3}.
A brief summary is given in the last section.

\section{Model and Method}\label{sec2}


We consider the following two-orbital Hubbard Hamiltonian,
\begin{eqnarray}
H&=&\sum_{\stackrel{<i,j>}{\alpha\beta\sigma}}
t_{ij}^{(\alpha\beta)} c_{i\alpha\sigma}^\dag c_{j\beta\sigma}
+U\sum_{i\alpha}n_{i\alpha\uparrow}n_{i\alpha\downarrow}\nonumber\\
&+&\left(U'-J\right)\sum_{i\sigma}n_{i1\sigma}n_{i2\sigma}
+U'\sum_{i\sigma}n_{i1\sigma}n_{i2\bar{\sigma}}\nonumber\\
&-&J\sum_{i}\left[ c_{i1\uparrow}^\dag c_{i1\downarrow}
c_{i2\downarrow}^\dag c_{i2\uparrow}+c_{i1\downarrow}^\dag c_{i1\uparrow}
c_{i2\uparrow}^\dag c_{i2\downarrow}\right] \nonumber\\
&-&J\sum_{i}\left[ c_{i1\uparrow}^\dag c_{i1\downarrow}^\dag
c_{i2\uparrow} c_{i2\downarrow}+ c_{i2\uparrow}^\dag c_{i2\downarrow}^\dag
c_{i1\uparrow} c_{i1\downarrow} \right]\label{eq:model}
\label{Hamilt}
\end{eqnarray}
where $c_{i\alpha\sigma}^\dag (c_{i\alpha\sigma})$ 
creates (annihilates) an electron 
with  spin $\sigma(=\uparrow, \downarrow)$ and orbital
index $\alpha(=1, 2)$ at the $i$th site and 
$n_{i\alpha\sigma}=c_{i\alpha\sigma}^\dag c_{i\alpha\sigma}$. 
$U$ ($U'$) represents the intraband (interband) Coulomb interaction and
$J$ the Hund coupling.
For electron hopping, we introduce
\begin{eqnarray}
t_{ij}^{(\alpha\beta)}=t_{ij}^{(\alpha)}\delta_{\alpha\beta}+V\delta_{ij},
\end{eqnarray}
with  the orbital-dependent nearest-neighbor hopping $t_{ij}^{(\alpha)}$ 
and the hybridization $V$ between two orbitals.
By this generalized model, we can study several different models 
in the same  framework. For $V=0$, 
the system is reduced to the multi-orbital Hubbard model 
with the same $(t_{ij}^{(\alpha)}=t_{ij})$
or distinct orbitals.\cite{KogaLett,Liebsch}
On the other hand, for $t_{ij}^{(2)}=0$, the system is reduced to 
a correlated electron system coupled to localized electrons, 
such as the periodic Anderson model ($J=0$) for 
heavy-fermion systems\cite{Coleman,Rice,Yamada,Kuramoto,Kim}
 or the double exchange model
($J>0$) for some transition metal oxides.
\cite{Zener,Anderson,Kubo,Furukawa}
For general choices of the parameters, we 
expect a variety of 
characteristic  properties inherent in these limiting models to appear naturally.


To investigate the above degenerate Hubbard model,
we make use of DMFT, \cite{Metzner,Muller,Georges,Pruschke}
which has successfully been applied to various electron systems such as 
the single band Hubbard model, 
\cite{Caffarel,2site,single1,OSakai,single2,single3,single4,BullaNRG}
the multi-orbital Hubbard model, 
\cite{2band1,2band2,Koga,Momoi,OnoED,Sakai,multi,sces,KogaLett,Liebsch}
the periodic Anderson model. \cite{PAM,Mutou,Saso,Sato,Ohashi,Medici}
In DMFT, the lattice model is mapped to
 an effective impurity  model, 
where local electron correlations are taken into account precisely. 
The lattice Green function is then obtained via self-consistent
conditions imposed on the impurity problem.

In  DMFT for the multi-orbital model,
the Green function in the lattice system is given as,
\begin{eqnarray}
{\bf G}\left(k, z\right)^{-1}={\bf G}_0\left(k, z\right)^{-1}
-{\bf \Sigma}\left(z \right),
\end{eqnarray}
with
\begin{equation}
{\bf G}_0\left( k, z\right)^{-1}=\left(
\begin{array}{cc}                 
z+\mu-\epsilon_1( k) & -V\\
-V & z+\mu-\epsilon_2( k)
\end{array}
\right),
\end{equation}
and
\begin{equation}
{\bf \Sigma}\left(z\right)=\left(
\begin{array}{cc}
\Sigma_{11}(z) & \Sigma_{12}(z) \\
\Sigma_{21}(z) & \Sigma_{22}(z) 
\end{array}
\right),
\end{equation}
where $\mu$ is the chemical potential, and $\epsilon_\alpha (k)$
is the bare dispersion relation 
for the $\alpha$-th orbital. 
In terms of the density of states $\rho (x)$ rescaled by the 
band width $D_\alpha$,
the local Green function is expressed as,
\begin{eqnarray}
G_{11}(z)&=&\int dx \frac{\rho(x)}{\xi_1\left(z,x\right)-
\frac{\displaystyle v(z)^2}{\displaystyle \xi_2\left(z, x\right)}},
\nonumber\\
G_{12}(z)&=&\int dx \frac{v(z)}
{\xi_1\left(z, x\right)\xi_2\left(z, x\right)-v(z)^2},
\nonumber\\
G_{22}(z)&=&\int dx \frac{\rho(x)}{\xi_2\left(z, x\right)-
\frac{\displaystyle v(z)^2}{\displaystyle \xi_1\left(z, x\right)}},
\end{eqnarray}
where
\begin{eqnarray}
 \xi_1\left(z, x\right)&=&z+\mu-\Sigma_{11}-D_1 x,\nonumber\\
 \xi_2\left(z, x\right)&=&z+\mu-\Sigma_{22}-D_2 x,\nonumber\\
 v\left(z\right)&=&V+\Sigma_{12}\left(z\right).
\end{eqnarray}
In the following, we use the semicircular density of states 
$\rho (x)=2/\pi \sqrt{1-x^2}$.


There are  various numerical methods 
to solve the effective impurity problem.
Note that the explicit model Hamiltonian for the impurity system 
is not obtained straightforwardly in our case,
since the lattice Green function has a frequency-dependent term in the 
non-diagonal element when the system has the hybridization $V$ and 
the finite band width in both orbitals. 
Therefore, it is not necessarily most efficient  to 
apply the exact diagonalization\cite{Caffarel} or
the two-site DMFT\cite{2site} methods
as impurity solvers, because these methods require the knowledge 
of the explicit form of the Hamiltonian. 
Furthermore, self-consistent perturbation theories such as
the iterative perturbation method and the non-crossing approximation
are not appropriate to discuss  orbital fluctuations
in the vicinity of the critical point.
In the present study, we make use of QMC
to treat the impurity model at finite temperatures.\cite{Hirsch}
In this connection, we note here that the Hund coupling plays
 a key role in controlling the nature of the Mott transition
in the multi-orbital system.\cite{sces}
Therefore, it is important to carefully analyze the effect of the Hund coupling 
in the framework of QMC.
To this end, we use the algorithm proposed by Sakai et al.,\cite{Sakai} 
where  the Hund coupling is represented in terms of
discrete auxiliary fields.
When we solve the effective impurity model by means of QMC method,
we use the Trotter time slices $\Delta \tau = (TL)^{-1} \le 1/6$,
where $T$ is the temperature and $L$ is the Trotter number.

In the following, we fix the band widths as $(D_1, D_2)=(1.0, 2.0)$ 
and the chemical potential as $\mu=-U/2-U'+J/4$ 
to discuss the metal-insulator transitions at half-filling.

\section{Results}\label{sec3}
\subsection{Non-hybridizing bands}

\begin{figure}[htb]
\begin{center}
\includegraphics[width=7cm]{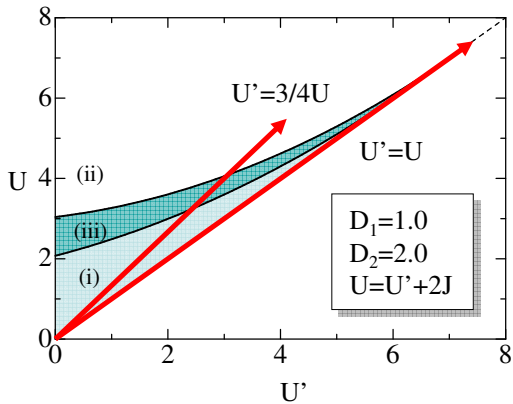}
\end{center}       
\caption{
Phase diagram for the two-orbital Hubbard model
with  $D_1=1$ and $D_2=2$. Note that the condition of rotational
symmetry,
$U=U'+2J$, is imposed (only the region of $U\ge U'$ is relevant).
 In the phase (i) (phase (ii)), 
both bands are  metallic (insulating), whereas
in the phase (iii)  
the metallic state coexists with the Mott insulating state.
Two lines along $U=U'$ with $J=0$  and
$U'/U=3/4$ with $J/U=1/8$ are shown, for which  
thermodynamic properties at finite temperatures are 
examined  in the text.
}
\label{fig:zero-phase}
\end{figure}

Before presenting the results computed at finite temperatures,
we briefly summarize the nature of the zero-temperature 
phase diagram for $V=0$ obtained by DMFT 
together with the exact diagonalization,\cite{KogaLett}  which 
is shown in  Fig. \ref{fig:zero-phase}.
There are three distinct phases depending on the strength of
the interactions.
It is seen that the metallic phase (i) remains stable up to  
large Coulomb interaction $U$ along the line
 $U \sim U'$ (small $J$), where
the Mott transitions merge to a single transition. 
Away from the symmetric limit, i.e.  $ U > U' $ 
with $ 2J = U - U' $,
 we find two separate Mott transitions in general.
In between the intermediate metallic phase (iii) appears
with one band localized and the other itinerant. 

We now analyze the temperature dependence of the charge, spin and 
orbital fluctuations 
 by combining DMFT with QMC simulations. We still restrict here
to the case of non-hybridized bands $(V=0)$.
Two typical sets of the parameters are considered, which satisfy
the conditions $(U'/U, J/U)=(3/4, 1/8)$ and $(1,0)$.  
As seen from Fig. \ref{fig:zero-phase},
the Mott transitions occur at two different critical points
$U_{c1}\sim 3$ and $U_{c2}\sim 4$ in the former case, while 
in the latter case they are merged to a single Mott transition 
at the critical point $U_c\sim 7$ for zero temperature.
The charge (c), spin (s) and orbital (o) susceptibilities are defined as 
\begin{eqnarray}
\chi_\gamma &=& \int_0^\beta 
d\tau \chi_\gamma(\tau),
\end{eqnarray}
with $\gamma=c, s, o$, and 
\begin{eqnarray}
\chi_c(\tau-\tau')&=& \langle
T|\left[n(\tau)-2\right]\left[n(\tau')-2\right]\rangle,\nonumber\\
\chi_s(\tau-\tau')&=& \langle
T|\left[n_\uparrow(\tau)-n_\downarrow(\tau)\right]
\left[n_\uparrow(\tau')-n_\downarrow(\tau')\right]\rangle,\nonumber\\
\chi_o(\tau-\tau')&=& \langle
T|\left[n_1(\tau)-n_2(\tau)\right]
\left[n_1(\tau')-n_2(\tau')\right]\rangle,
\end{eqnarray}
where $T$ is the time-ordering operator, 
$n(\tau)=\sum_{\alpha\sigma} n_{\alpha\sigma}(\tau)$, 
$n_\alpha(\tau)=\sum_{\sigma} n_{\alpha\sigma}(\tau)$,
$n_\sigma(\tau)=\sum_{\alpha} n_{\alpha\sigma}(\tau)$, and
$\tau$ is an imaginary time.

\begin{figure}[htb]
\begin{center}
\includegraphics[width=7cm]{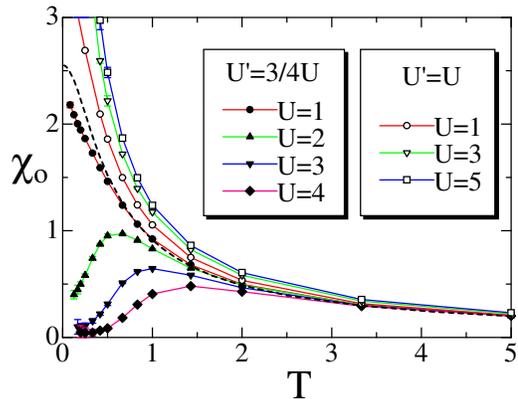}
\end{center}       
\caption{
Orbital susceptibility as a function of the temperature $T$ for $V=0$. 
Open (solid) symbols represent the results in the case $U=U'$ and $J=0$ 
($U'/U=3/4$ and $J/U=1/8$) and dashed lines those for the 
non-interacting case.
}
\label{fig:chi-o}
\end{figure}
We first turn to the orbital fluctuations. The temperature-dependent 
orbital susceptibility is shown in Fig. \ref{fig:chi-o}.
In the non-interacting system,
the orbital susceptibility increases with decreasing temperature,
and reaches a constant value at zero temperature.
If we now turn on the interactions (fixing  
the ratios $U'/U=3/4$ and $J/U=1/8$), the orbital susceptibility is suppressed
at low temperatures.  
This implies that electrons tend to localized in each band 
independently such that onsite fluctuations are unfavorable.  Eventually 
for $U \ge U_{c1} \sim 3$, one of the orbitals is entirely localized, 
so that orbital fluctuations are suppressed completely, giving 
$\chi_o=0$ at $T=0$.

On the other hand, very different behavior can be seen 
along the line $U'=U$ in Fig. \ref{fig:zero-phase}.
In this case, the
orbital susceptibility is increased with growing interactions
even at low temperatures. 
Interpreting this result in the context of the  phase diagram in Fig. \ref{fig:zero-phase},
we can say that the enhanced orbital fluctuations are 
relevant for stabilizing the metallic phase in the
strong correlation regime.
While such behavior is naturally expected
for models with
two equivalent orbitals, it appears even in systems with
nonequivalent bands.\cite{Koga}

\begin{figure}[htb]
\begin{center}
\includegraphics[width=8cm]{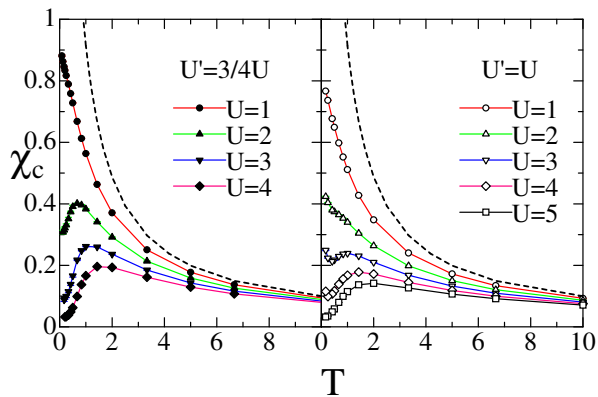}
\end{center}       
\caption{
Charge susceptibility as a function of the temperature $T$ for $V=0$.
Dashed lines represent the results for the non-interacting case.}
\label{fig:chi-c}
\end{figure}
To examine whether the system
shows metallic or insulating properties at finite temperatures,
we calculate the charge susceptibility (compressibility).
The obtained results are shown in Fig. \ref{fig:chi-c}.
In the case $U'/U=3/4$ and $J/U=1/8$,  the system with $U=3$
is located near the critical point between the metallic phase 
(i) and the intermediate phase (iii).
With decreasing temperature the 
charge susceptibility  decreases  below $T \sim 1$.
The appearance of a pseudogap  feature in an intermediate
temperature range gives rise to a depletion of density of
states at the Fermi energy for both bands.
Upon further lowering of the temperature
the charge susceptibility converges to a finite 
value, since the system still remains in a 
metallic phase, at least for one of the two orbitals.
For $U=4$, which corresponds to the boundary between 
the phases  (ii) and (iii), the charge susceptibility at low 
temperatures is almost zero, suggesting that the system has become
completely insulating corresponding to phase (ii). 
In contrast for $U'=U$ we observe in an intermediate range of
$U $ that with lowering temperature a decrease of the charge susceptibility
is followed by an eventual increase at lowest temperatures (Fig. \ref{fig:chi-c}).
Comparing this with Fig. \ref{fig:chi-o}, we see that the enhanced orbital fluctuations 
indeed have a tendency to stabilize  the metallic state.

We now  move to the spin susceptibility. 
\begin{figure}[htb]
\begin{center}
\includegraphics[width=8cm]{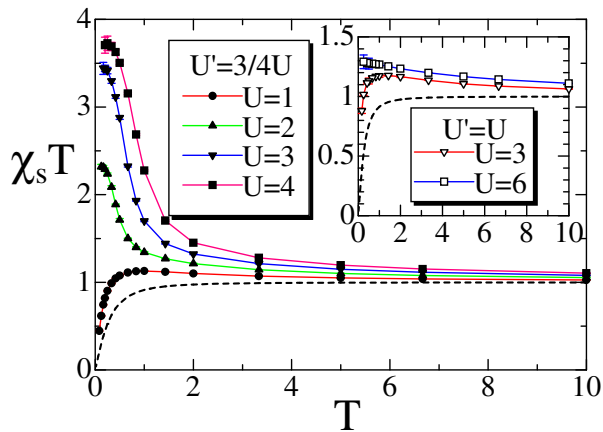}
\end{center}       
\caption{
The effective Curie constant
 $\chi_s T$ as a function of the temperatures $T$ 
for $U'/U=3/4$ and 
$J/U=1/8$. Inset shows the results in the case $U'=U$ and $J=0$.
Dashed lines represent the results for the non-interacting case.
}
\label{fig:chit-s}
\end{figure}
In Fig. \ref{fig:chit-s},  we plot 
the effective Curie constant  $\chi_s T$  as a function 
of the temperature.
We first look at  the case of $U'/U=3/4$ and $J/U=1/8$.
At high temperatures, all the spin configurations are equally 
populated, so that the effective Curie constant takes the value
$1/2$ for each orbital in our units, yielding $\chi_s T\sim 1$.
When electron correlations are weak
($U=1$), the system is still in the metallic phase,
so that the Pauli paramagnetic behavior with a constant 
 $\chi_s$ emerges, leading to $\chi_s T \rightarrow 0$  
as $T \rightarrow 0$. It is seen that the increase of the interactions
enhances the spin susceptibility at low temperatures, as a result of
the progressive trend to localize the electrons. 
The effective Curie constant is $\chi_sT=2$ when a free spin is realized
in each orbital.
It is seen that the Curie constant increases beyond the value of 
2 with the increase of the interactions ($U=3, 4$).
This means that ferromagnetic correlations due to 
the Hund coupling appear here. 

When $U'=U$ (inset of Fig. \ref{fig:chit-s}), both  spin
and orbital fluctuations are enhanced in the 
presence of the interactions. 
Accordingly, both  spin and orbital susceptibilities
increase at low temperatures, forming  heavy-fermion states
as far as the system stays in the metallic phase 
(see also Fig. \ref{fig:chi-o}).  Note that for $U=6$, at which 
the system is  close to the  Mott
transition point, the spin susceptibility is enhanced with the effective
Curie constant $\chi_sT \sim 4/3$ down to very low temperatures,
as seen in the inset of Fig. \ref{fig:chit-s}.
The value of 4/3 immediately follows if one takes into account 
two additional  configurations of doubly-occupied orbital besides
 four magnetic configurations, which are all degenerate
 at the metal-insulator 
transition point. Although not clearly observed
in the temperature range shown, $\chi_sT$ should vanish at
zero temperature for $U=U'=6$, since the system is still in 
the metallic phase, as seen from Fig. \ref{fig:zero-phase}.

\begin{figure}
\begin{center}
\includegraphics[width=8cm]{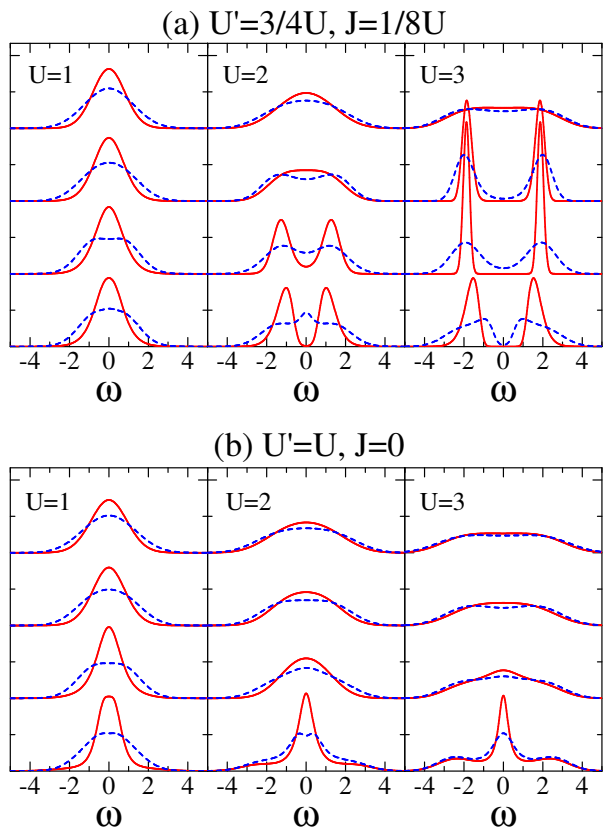}
\end{center}       
\caption{Density of states for the degenerate Hubbard model $(D_1,
 D_2)=(1.0, 2.0)$. The data are for the temperatures $T=2, 1,
 1/2$ and $1/6$ from the top to the bottom.
}
\label{fig:dos0}
\end{figure}

To see the above characteristic properties more clearly,
 we show 
the density of states for each orbital in Fig. \ref{fig:dos0},
which is computed by the maximum entropy method.\cite{MEM1,MEM2,MEM3}
When the interactions increase along the line  $U'/U=3/4$ and $J/U=1/8$,
 the OSMT should occur. Such tendency indeed 
appears at low temperatures in Fig. \ref{fig:dos0}(a).
Although both orbitals stay in metallic states down to
low temperatures ($T=1/6$) for $U=1$,  the OSMT
seems to occur for $U=2$; namely one of the bands develops the Mott Hubbard
gap, while the other band still remains metallic.
At a first glance, this result is slightly different from
 the ground-state phase diagram 
shown in Fig. \ref{fig:zero-phase}, where the system is in 
the phase (i) even at  $U=2$.  
However, this deviation is naturally understood
if we take into account the fact that for $U=2$, the 
narrower band is already in a highly correlated  
 metallic state, so that the sharp quasi-particle peak immediately 
disappears as the temperature increases beyond 
the small characteristic energy scale. This explains the behavior 
observed in the density of states at $T=1/6$.  For $U=3$, both 
bands are insulating at $T=1/6$ (the system
is almost on the boundary between the phases (ii) and (iii) at 
zero temperature).

In the case $U'=U$, as expected we encounter the qualitatively different 
behavior shown in Fig. \ref{fig:dos0}.
 In this case, 
both bands gradually develop quasi-particle peaks
 as the interactions increase, and 
 they still remain metallic even at $U=U'=3$.
  As mentioned above, all these features which are in contrast 
to  the situation for $U' \neq U$, 
are caused by the special symmetry for  $U=U'$, which gives rise to
equally enhanced spin and orbital fluctuations.

\subsection{Hybridization between distinct orbitals}

We have so far treated the degenerate Hubbard model, in which 
two types of orbitals do not mix with each other.
In our treatment with DMFT, the Mott insulating phase (ii) as well as 
the intermediate phase (iii) may be unstable 
against certain perturbations. There may be several possible 
mechanisms that stabilize such insulating phases.
One of the mechanisms,
which may play an important role in real materials,
is the hybridization between
the two distinct orbitals. We address the effect in this section.

This hybridization effect is relevant in some real materials.
For instance, 
in the compound $\rm Ca_{2-x}Sr_x Ru O_4$, \cite{Nakatsuji}
the hybridization between $\{\alpha, \beta\}$ and $\gamma$ orbitals 
is induced by the tilting of RuO$_6$ octahedra in the 
region of  $\rm Ca$-doping $0.2<x<0.5$,\cite{tilting}. This leads to 
Kondo-lattice like effective model and 
may be connected
with the reported heavy fermion behavior, 
\cite{Nakatsuji} similar to some $f$-electron systems.
This interesting aspect motivates us to study  the mixing effect
between the localized and itinerant electrons in the 
intermediate phase (iii).  Moreover
the compound $\rm La_{n+1}Ni_nO_{3n+1}$\cite{LaNiO} possesses
hybridization between $d_{3z^2-r^2}$ and $d_{x^2-y^2}$ orbitals
in the $e_g$ subshell. The OSMT may lead to the metallic but the 
less-conducting state is realized
below the critical temperature $T_c=550K$.\cite{Kobayashi}
Consequently we would like also to explore how the hybridization 
of different-type $d$-bands affects 
electronic properties especially around
the OSMT.

\begin{figure}
\begin{center}
\includegraphics[width=8cm]{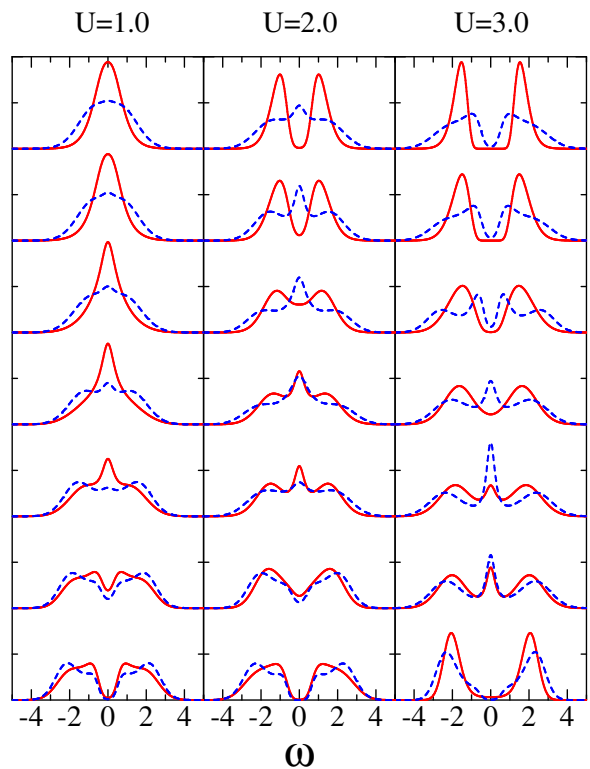}
\end{center}       
\caption{Solid (dashed) lines represent the density of states 
for the orbital $\alpha=1$ ($\alpha=2$)
when $(D_1, D_2)=(1.0, 2.0)$ at $T=1/6$
with  the  fixed  parameters of $U'/U=3/4$ and $J/U=1/8$.
The data are plotted for $V=0.0, 0.25, 0.5, 0.75, 1.0, 1.25$ 
and $1.5$ from top to bottom.
}
\label{fig:dos-V}
\end{figure}

We study the general case with
$U'\neq U$ and $J\neq 0$ in the presence of the hybridization
$V$. In Fig. \ref{fig:dos-V}, the 
 density of states calculated by the maximum entropy
method is shown for different choices of  $V$.
We start with the weak coupling case, $U=1$, where
the metallic states are realized in both orbitals at $V=0$.
Although the introduction of small $V$ does not alter
the nature of the ground state, 
further increase of $V$ splits
the density of states ($V=1.5$), signaling the formation of the 
band insulator: namely all kinds of elementary 
excitations possess the gap.
In contrast, we encounter different behavior when electron
interactions are increased up to $U=2$ and 3. In these parameters,
 the system at $V=0$ shows the intermediate
or Mott-insulating properties at $T=1/6$. It is seen that
the density of states around the Fermi level increases
as $V$ increases.  For $U=2$, the intermediate state
is first changed to a metallic state, where the quasi-particle 
peaks appear in both orbitals ($V=0.75,1.0$). 
For fairly large $V$, both bands fall into the 
renormalized band  insulator ($V=1.5$).
Similarly, for $U=3$, the hybridization first drives the Mott-insulating
state to an intermediate one, as is clearly seen at $V=0.75$, which 
is followed by two successive transitions as is the case for $U=2$.

\begin{figure}
\begin{center}
\includegraphics[width=8cm]{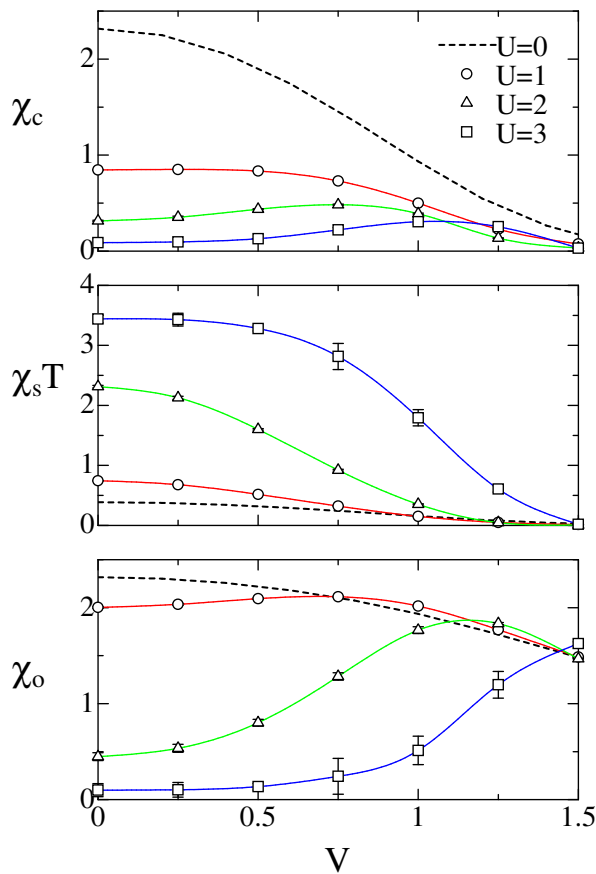}
\end{center}       
\caption{Charge, spin and orbital susceptibilities 
as a function of the hybridization $V$ at the temperature $T=1/6$.
}
\label{fig:chiall-v}
\end{figure}

The above characteristic properties also emerge in the
charge, spin and orbital susceptibilities
at low temperature, as shown in Fig. \ref{fig:chiall-v}.
For weak interactions ($U=1$), the charge 
susceptibility $\chi_c$ monotonically decreases with the increase of $V$.
When electron correlations become strong, 
the non-monotonic behavior appears in $\chi_c$:
the charge fluctuations, which are suppressed at $V=0$,
are somewhat recovered by the hybridization, 
which leads to metallic behavior. 
For large $V$, $\chi_c$ is again  suppressed
since the system turns into a band insulator.
We can see that the orbital susceptibility exhibits 
non-monotonic behavior similar to the charge susceptibility,
 the origin of which is essentially the
same as in $\chi_c$; the orbital fluctuations suppressed 
at $V=0$ are recovered by $V$, and then 
the formation of  the band insulator causes 
the gradual decrease of $\chi_o$.
In contrast,  the spin susceptibility  monotonically 
decreases with the increase of  $V$ irrespective of 
the strength of the interactions.  As discussed for
$V=0$, the effective spin is enhanced by ferromagnetic 
fluctuations due to the Hund coupling in the insulating 
and intermediate phases. Upon introducing the hybridization in 
these phases, the ferromagnetic fluctuations are 
suppressed, leading to the monotonic decrease of the 
effective Curie constant.

\begin{figure}
\begin{center}
\includegraphics[width=8cm]{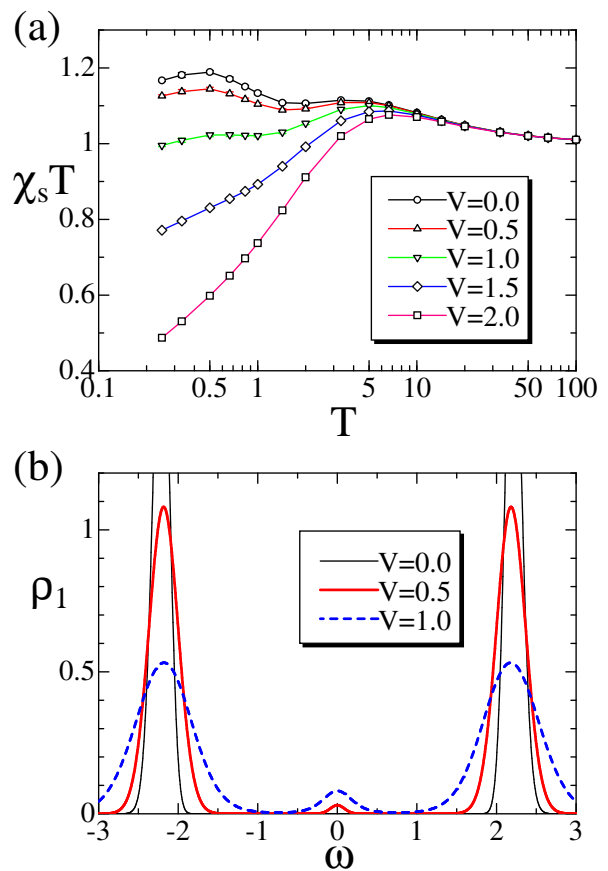}
\end{center}      
\caption{(a) Effective Curie constant as a function of the temperature 
and (b) density of states in the narrower band $(\alpha=1)$  
at $T=1/4$  for
an extreme choice of the bandwidths, $(D_1,D_2)=(1.0, 10.0)$. 
The density of states for the wider band is not shown here.
The other parameters are $U=4.0, U'=3.0$ and $J=0.5$.
}
\label{fig:ex}
\end{figure}

 From the above observations, we can say that the introduction of 
appropriate hybridization induces 
heavy-fermion metallic behavior.
In fact, this tendency can be observed more clearly in an
extreme choice of the bandwidths, $(D_1, D_2)=(1.0, 10.0)$,
 shown in Fig. \ref{fig:ex}. 
At $V=0.0$, the system is in the intermediate phase, so that
the completely localized states [Fig. \ref{fig:ex} (b)] appear
 in the narrower band in the background of 
the nearly free bands. This double structure in the system gives rise to
two peaks in the temperature-dependent effective Curie constant,
as shown in Fig. \ref{fig:ex} (a).
Since the completely localized state plays a role of the
$f$-state in the Anderson lattice model,
\cite{Kusunose} a ''heavy-fermion'' peak appears
at the Fermi energy in the 
presence of $V$, which is essentially the 
same as  that observed in Fig. \ref{fig:dos-V}.

%

\begin{figure}
\begin{center}
\includegraphics[width=8cm]{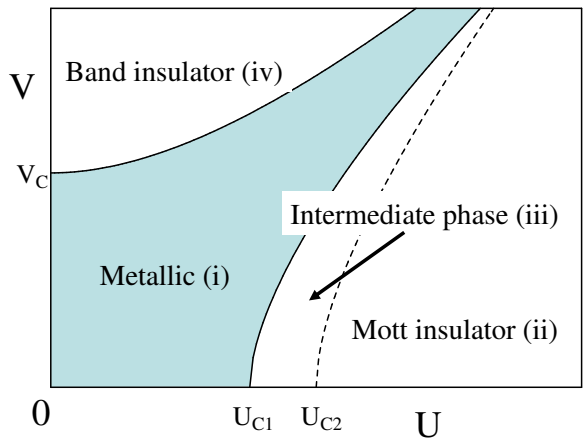}
\end{center}       
\caption{
Schematic phase diagram for the two-orbital  Hubbard model 
with finite hybridization between two orbitals.
Solid lines represent the phase boundaries between the metallic and 
insulating phases. 
Dashed line indicates the crossover between the Mott insulator and 
the Kondo insulator.
}
\label{fig:phase}
\end{figure}

Finally, some comments are in order on the 
phase diagram at zero temperature.
In our approach, it is not easy to deal with the system 
at very low temperatures, since  QMC simulations  suffer
 from minus sign problems.
Nevertheless, we may give some qualitative arguments on the 
expected phase diagram at zero temperature.
As discussed above, the metallic phase (i) is not so sensitive to 
$V$ as far as it is small. This is also the case for the completely 
insulating phase (ii). In contrast, a more subtle situation appears
in the intermediate phase (iii).  As mentioned above,
the intermediate phase exhibits Kondo-like heavy fermion
behavior at low temperatures in the presence of $V$. 
Recall, however,  that we are now concerned 
with the half-filled band. Therefore, this Kondo-like metallic phase 
should acquire a Kondo-insulating gap due to commensurability
at zero temperature.  We would thus say that the intermediate
phase (iii) is changed into the Kondo-insulator with a tiny
excitation gap in the presence of $V$ at zero temperature. Accordingly, 
 the  sharp transition between the phases (ii) and (iii) at $V=0$
may  be smeared and changed to crossover behavior.
These considerations lead us to 
a schematic description of phase diagram for the two-orbital model
with mixing between the 
distinct orbitals, as shown in Fig. \ref{fig:phase}.
On the line of $V=0$, the OSMT,
which may occur in general choices of the parameters, separates 
the phase at $V=0$ into three regions.
The metallic phase for small $U$ is simply driven to 
the band-insulator (iv) beyond  a certain critical value of hybridization.
The intermediate phase  at $V=0$ is 
 changed to the Kondo-insulator 
in the presence of any finite $V$. This insulating state 
first undergoes a phase transition to the metallic phase, and
eventually enters the band-insulator as $V$ increases.
The completely Mott insulating phase first shows a crossover to
the Kondo insulator, which is further driven to the metallic phase 
and then to the band-insulating phase.
Note that at finite temperatures above the Kondo-insulating gap,
we can observe  a Kondo-type heavy
fermion behavior in the intermediate phase  with finite $V$.


\section{Summary}

We have investigated the degenerate Hubbard model with distinct 
hopping integrals by combining DMFT with 
QMC simulations. By examining the spin, charge 
and orbital susceptibilities calculated at finite temperatures,
we have clarified that equally enhanced spin and 
orbital fluctuations play a vital role
on stabilizing the metallic states in the multi-orbital systems.
This remarkable effect is responsible for whether the system
undergoes a single Mott transition or OSMTs.
Also,  we have discussed how the phase diagram at
finite temperatures slightly
deviates from the ground-state one because of smearing 
effect of the narrow quasi-particle peak.

We have further explored the effect of the hybridization between 
the distinct orbitals, and have found that it plays a
crucial role especially around  the OSMT.
The introduction of the hybridization in  the intermediate phase 
 enhances the charge and 
orbital fluctuations,  inducing the metallic phase 
with a sharp quasi-particle peak.
Accordingly,  Kondo-like heavy fermion states show up 
at finite temperatures, which eventually drop in
the Kondo insulating phase for our half-filled bands.
We have also pointed out that the hybridization effect  smears the sharp 
OSMT at zero temperature, and changes it to a
crossover behavior.  Nevertheless, we can still observe 
the OSMT at finite temperatures.

In this paper, we have used QMC as an impurity solver
in DMFT, which is not powerful enough to treat  properties at 
very low temperatures. Therefore, it is desirable to 
exploit a complementary approach to study such low-temperature 
properties more precisely, although we have arrived at a reasonable 
phase diagram at zero temperature.  Various remaining open
problems could not be addressed in the present study.
 One of the most important issues to explore is
 magnetism of the system, which has not been seen here, 
since we have restricted our attention to the 
paramagnetic phase. This problem is under 
consideration.

\section{Acknowledgments}
We would like to thank K. Ishida, S. Sakai, S. Nakatsuji and Y. Maeno 
for useful discussions.
This work was partly supported by a Grant-in-Aid from the Ministry 
of Education, Science, Sports and Culture of Japan, 
the Swiss National Science Foundation and the Centre of Theoretical Studies
at ETH Z\"urich.
A part of computations was done at the Supercomputer Center at the 
Institute for Solid State Physics, University of Tokyo
and Yukawa Institute Computer Facility.

%


\end{document}